\documentclass[twocolumn,showpacs,prb,citeautoscript,amsmath,amssymb,floatfix]{revtex4-1}
\usepackage{amsmath,amssymb,color}
\usepackage{bm}
\usepackage{graphicx}
\usepackage{psfrag}
\usepackage{hyperref}
\newcommand{\bd}{\bm}

\begin{document}

\title{Spin currents, spin torques, and the concept of spin superfluidity}

\author{Andreas R\"{u}ckriegel and Peter Kopietz}
  
\affiliation{Institut f\"{u}r Theoretische Physik, Universit\"{a}t
  Frankfurt,  Max-von-Laue Strasse 1, 60438 Frankfurt, Germany}

\date{January 19, 2017}

\begin{abstract}
In magnets with non-collinear spin configuration 
the expectation value of the conventionally defined spin current operator contains a contribution
which renormalizes an external magnetic field
and hence affects only the  precessional motion of the spin polarization. 
This term, which has been named angular spin current by
Sun and Xie [Phys. Rev B \textbf{72}, 245305 (2005)], 
does not describe the translational motion of magnetic moments.
We give a prescription how to separate these two types of spin transport
and show that the translational movement of the spin is always polarized 
along the direction of the local magnetization.
We also show that at vanishing temperature the classical magnetic order parameter in magnetic insulators
cannot carry a translational spin current,
and elucidate how this affects the interpretation of spin supercurrents.
\end{abstract}


\maketitle

\section{Introduction}


The notion of  spin currents describing the motion of magnetic moments associated with the spins of the electrons in solids is  of central importance in the field of spintronics where
one tries to use the spin degree of freedom to store and
process information. Unfortunately, in systems lacking spin-rotational invariance
(which can be broken by an external magnetic field or by relativistic effects
such as spin-orbit coupling or dipole-dipole interactions)
the proper definition of the quantum mechanical operator representing the spin current
is ambiguous, because the magnetization does not satisfy a local conservation law.
In the past decade several authors have proposed
resolutions of this ambiguity 
\cite{Rashba03,Rashba04,Rashba05,Schuetz03,Schuetz04,Culcer04,Bruno05,Sun05,Shi06,Zhang08,Bostrem08,Tokatly08,Bray09,An12,Berche12,Nakata14}, but a generally accepted
agreement on the correct definition of the spin current operator in systems
without spin conservation has not been found.

\begin{figure}[tb]
\begin{center}
 \includegraphics[width=0.4\textwidth]{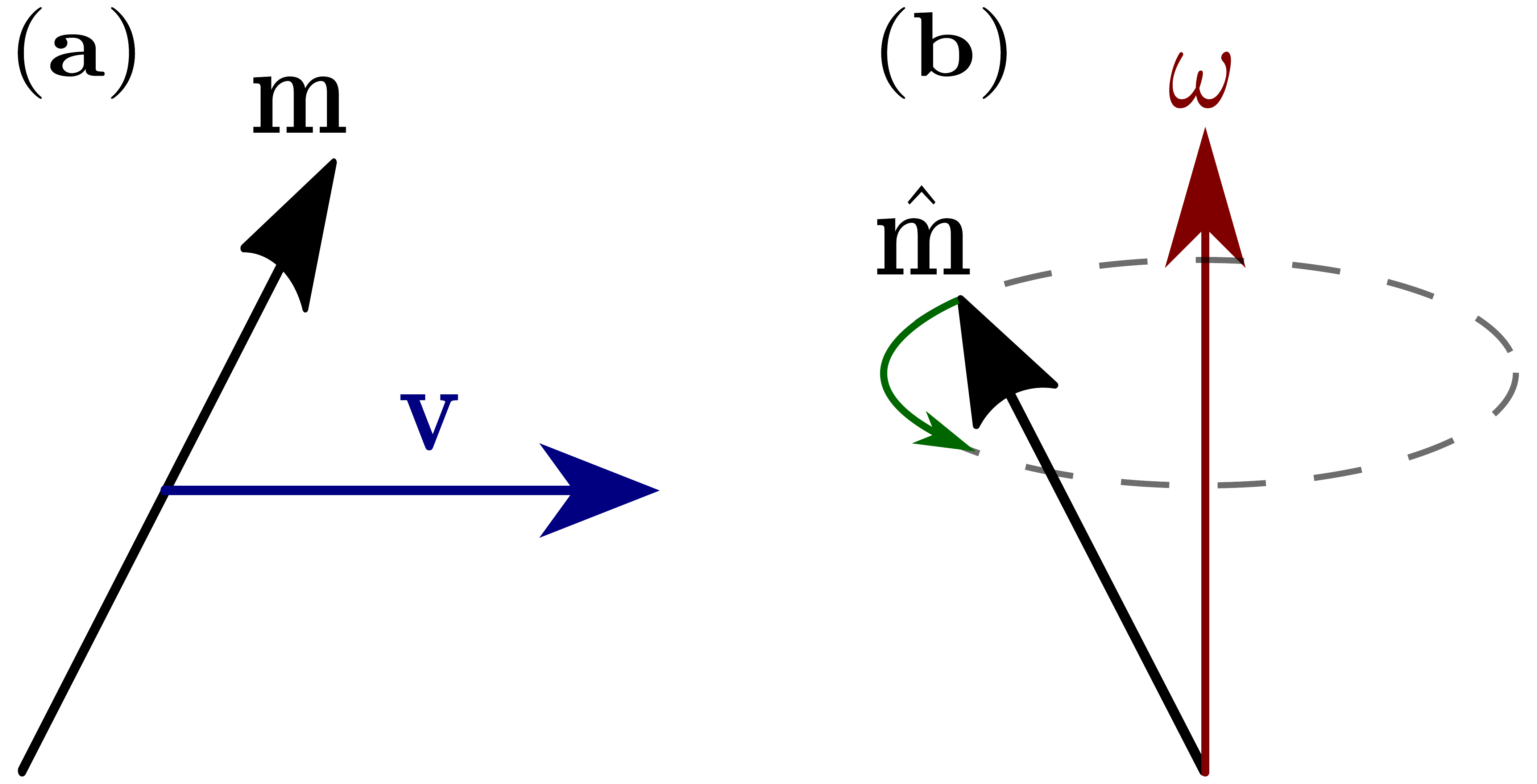}
\end{center}
\caption{%
(Color online)
The two kinds of spin transport.
(a) Translational motion of the magnetic moment $\bm{m}$ with a velocity $\bm{v}$,
corresponding to the physical movement of a magnetic moment with magnitude $|\bm{m}|$. 
(b) Precessional motion of the magnetic polarization $\hat{\bm{m}}=\bm{m}/|\bm{m}|$ 
with a frequency $\bm{\omega}$.
The latter motion is purely angular and leaves the magnitude $|\bm{m}|$ of the magnetic moment invariant.
}
\label{fig:transport}
\end{figure}

The purpose of this work is show that the distinction between
translational and angular spin currents proposed by
Sun and Xie \cite{Sun05} leads to a simple and unique definition of the concept of
spin transport in condensed matter systems.
Sun and Xie \cite{Sun05} pointed out 
that spin currents describe moving magnetic dipoles, and that 
generally the transport of any vector
can be decomposed into a translational part characterized by some velocity
$\bd{v} ( \bd{r} )$ and an angular part described by some angular velocity
$\bd{\omega} ( \bd{r} )$,
see Fig.~\ref{fig:transport}.
In the context of spin transport
Sun and Xie \cite{Sun05} called the latter contribution the
{\it{angular spin current}}, although this can be also viewed as the spin torque
discussed earlier by
Culcer {\it{et al.}}~[\onlinecite{Culcer04}].
We will explicitly show below that the equation of motion of a magnetic moment $\bm{m}_i$
at lattice site $\bm{R}_i$
with magnitude $m_i=|\bm{m}_i|$ 
and polarization $\hat{\bm{m}}_i=\bm{m}_i/|\bm{m}_i|$ 
can be decomposed into a translational part
\begin{equation} \label{eq:translation}
\partial_ t m_i + \sum_j \langle I^{\parallel}_{i\to j} \rangle = 0 ,
\end{equation}
that corresponds to the physical movement of magnetic dipole moments,
and an angular part
\begin{equation} \label{eq:polarization}
\partial_ t \hat{\bm{m}}_i = \bm{\omega}_i \times \hat{\bm{m}}_i ,
\end{equation}
that describes the precessional motion of the magnetization.
We will provide expressions for
the translational spin current operator $I^{\parallel}_{i\to j}$
and the precession frequency $\bm{\omega}_i$
valid for itinerant as well as localized systems.
Although the explicit construction of this decomposition is very simple, 
it entails profound physical consequences:
Since only the translational movement of magnetization corresponds to
the physical displacement of magnetic moments,
in equilibrium only stationary translational spin currents can generate an electrical field \cite{Hirsch99}.
Angular spin currents on the other hand only transport the polarization,
hence a stationary angular spin current is simply an inhomogeneous configuration of the local magnetic order and does not create an electrical field.
Furthermore, we also show that in generic magnetic insulators the classical magnetic order cannot support translational spin transport at vanishing temperature;
incoherent thermal or quantum fluctuations are mandatory for the physical transport of magnetization in these systems.
This also implies that spin superfluidity in magnetic insulators \cite{MacDonald01,Nogueira04,Sonin10,Sonin13,MacDonald14,Takei14a,Takei14b,Flebus16,Chen16} must be angular spin transport
that can be visualized as transporting the spin polarization,
but does not correspond to the physical movement of magnetic moments.

The remainder of this work is organized as follows:
In Sec.~\ref{sec:Separation} we will derive general expressions 
for the operators corresponding to the translational spin current
and to the precession frequency of the magnetization,
first for itinerant systems and then for localized magnetic moments.
We proceed to illustrate the usefulness of the decoupling procedure in Sec.~\ref{sec:supercurrents},
where we discuss spin superfluidity in easy-plane ferromagnets
and persistent spin currents in Heisenberg rings.
Finally, in Sec.~\ref{sec:conclusions} we present our conclusions.
The Appendix contains some additional details of the self-consistent spin-wave expansion 
we employ to describe easy-plane ferromagnets.

\section{Separating translational from angular spin transport}
\label{sec:Separation}

In this section, we explicitly show how translational and
angular spin transport can be defined on the operator level.

\subsection{Itinerant electrons}

To construct the proper quantum mechanical 
definition of the translational spin transport operator
let us consider a lattice model describing electrons with 
spin-dependent hopping 
$t_{ij}^{\sigma \sigma^{\prime}}$
in an inhomogeneous magnetic field $\bd{h}_i$.
The second quantized Hamiltonian of our model is
 \begin{equation}
 {\cal{H}}   = \sum_{ij \sigma \sigma^{\prime}} t_{ij}^{\sigma \sigma^{\prime}} 
 c^{\dagger}_{ i \sigma} c_{j \sigma^{\prime}}   - \sum_i \bd{h}_i \cdot \bd{s}_i    +  
{\cal{U}},
 \label{eq:Hubbard}
 \end{equation}
where ${\cal{U}}$ is some spin-rotationally invariant interaction,
$c_{i \sigma}$ annihilates a fermion with spin-projection $\sigma$ at lattice site
$\bd{R}_i$, and
the itinerant spin operators are defined by
 \begin{equation}
 \bd{s}_i =   \frac{1}{2} {c}^{\dagger}_i \bd{\sigma} {c}_i, \; \; \; 
 {c}_i = \left( \begin{array}{c} c_{i \uparrow} \\ c_{i \downarrow} 
 \end{array} \right).
 \end{equation}
Here $\bd{\sigma}$ is the vector of Pauli matrices.
The spin-dependent hopping energies $t_{ij}^{\sigma \sigma^{\prime}}$ are of the form
 \begin{equation}
t_{ij}^{\sigma \sigma^{\prime}} = t_{ij} \delta_{\sigma \sigma^{\prime}} + i ( \bd{\lambda}_{ij} \cdot
 \bd{\sigma} )_{\sigma \sigma^{\prime} } ,
 \end{equation}
where  the vectors $\bd{\lambda}_{ij}$ 
are proportional to the strength of the spin-orbit coupling.
The hermiticity of the Hamiltonian implies the symmetries
$t_{ij} = t^{\ast}_{ji}$ and $\bd{\lambda}_{ij} = - \bd{\lambda}^{\ast}_{ji}$.
Using the canonical anticommutation relations $ \{ c_{i \sigma} , c^{\dagger}_{j \sigma^{\prime}}
 \} = \delta_{ij} \delta_{\sigma \sigma^{\prime}}$ and the fact that the interaction is  spin-rotationally
invariant,
$ [ \bd{s}_i , {\cal{U}} ] =0$,
we obtain the Heisenberg equation of motion for the itinerant spins,
 \begin{equation} 
 \frac{ \partial \bd{s}_i }{\partial t } 
 + \sum_j   \bd{I}_{i \rightarrow j}
 = -  \bd{h}_i \times \bd{s}_i  ,
 \label{eq:eomit}
 \end{equation}
where we have defined the operator
 \begin{eqnarray}
  \bd{I}_{i \rightarrow j} & = & 
- \frac{1}{2i}
 \left( t_{i j} {c}^{\dagger}_i \bd{\sigma} {c}_j - t_{ij}^{\ast}
 {c}^{\dagger}_j \bd{\sigma} {c}_i \right)
 \nonumber
 \\
 &   & - \frac{1}{2}  \left( \bd{\lambda}_{ij}    c^{\dagger}_i c_j  
  + \bd{\lambda}_{ij}^{\ast}   c^{\dagger}_j c_i \right)
 \nonumber
 \\ 
&   & -
\frac{1}{2i}
 \left( {c}^{\dagger}_i ( \bd{\sigma} \times \bd{\lambda}_{ij} ) {c}_j -
 {c}^{\dagger}_j (  \bd{\sigma}  \times  \bd{\lambda}_{ij}^{\ast}  ) {c}_i 
  \right).
 \label{eq:Islattice}
 \end{eqnarray}
It is tempting to associate  this operator 
with the spin current describing the transport of spin from  lattice site $\bd{R}_i$ to lattice site 
$\bd{R}_j$.  It turns out, however, that a certain part of this operator simply renormalizes 
the external magnetic field and therefore cannot be associated with translational 
spin transport.
To isolate this contribution and identify the angular part which renormalizes the
precessional motion of the spins, we take
the  quantum mechanical expectation value
of both sides of the equation of motion (\ref{eq:eomit}) and 
obtain a formally exact equation of motion for the magnetic
moments $\bd{m}_i ( t ) = \langle \bd{s}_i ( t ) \rangle$,
 \begin{equation}
 \partial_t \bd{m}_i  + \bd{T}_i = - \bd{h}_i \times
 \bd{m}_i,
 \label{eq:eomD}
 \end{equation}
where the spin torque is defined by
 \begin{equation}
 \bd{T}_i = \sum_{j} \langle  \bd{I}_{i \rightarrow j}   \rangle.
 \end{equation}
To identify the contribution responsible for translational spin transport, 
we further decompose the vector  $\bd{T}_i$
into a longitudinal and a transverse part,
 \begin{equation}
 {\bd{T}}_i = T_i^{\parallel} \hat{\bd{m}}_i + \bd{T}_i^{\bot},
 \end{equation}
where $\hat{\bd{m}}_i = \bd{m}_i / | \bd{m}_i| $ 
is the local spin polarization and
 \begin{eqnarray}
 T_i^{\parallel} & = & \sum_j \hat{\bd{m}}_i \cdot \langle  \bd{I}_{i \rightarrow j}  \rangle,
 \\
 \bd{T}_i^{\bot} & = & \bd{T}_i -  (  
\bd{T}_i   \cdot \hat{\bd{m}}_i  ) \hat{\bd{m}}_i.
 \end{eqnarray}
Writing
 \begin{equation}
  \bd{T}_i^{\bot} = 
  ( \hat{\bd{m}}_i \times \bd{T}_i ) \times \hat{\bd{m}}_i 
 = \delta \bd{h}_i^{\bot} \times \bd{m}_i,
 \end{equation}
where
 \begin{equation}
 \delta \bd{h}_i^{\bot} = \frac{ \hat{\bd{m}}_i \times \bd{T}_i }{ | \bd{m}_i | }
 =  \sum_{j} \frac{\hat{\bd{m}}_i}{ | \bd{m}_i | } \times
 \langle \bd{I}_{ i \rightarrow j } \rangle
 \end{equation}
is the induced magnetic field perpendicular to the direction of $\bd{m}_i$,
we see that the transverse part $\bd{T}_i^{\bot}$
renormalizes the external magnetic field.
The total angular frequency relevant for the precessional motion 
of the magnetic moments is
 \begin{equation}
 \bd{\omega}_i = - \bd{h}_i -  \delta \bd{h}^{\bot}_i.
 \end{equation}
The term $- \delta {\bd{h}}_i^{\bot} \times \bd{m}_i$ can be called 
angular spin current \cite{Sun05} or spin torque \cite{Culcer04} and should 
be added to the external
torque $- \bd{h}_i \times \bd{m}_i$ acting on the magnetic moments.
The expectation value 
of the  equation of motion  (\ref{eq:eomD}) can now be written as
  \begin{equation}
 \partial_t \bd{m}_i +  \hat{\bd{m}}_i   T_i^{\parallel}
 =  \bd{\omega}_i \times
 \bd{m}_i.
 \label{eq:eommagomega}
 \end{equation}
From this expression it is easy to show that the spin torque does not contribute
to the time-evolution of the magnitude
$m_i = | \bd{m}_i |$ of the magnetic moments, which satisfies the equation of 
motion\cite{Schuetz04}
 \begin{equation}
 \partial_t m_i +
\sum_{j} \hat{\bd{m}}_i \cdot \langle  \bd{I}_{i \rightarrow j}  \rangle =0.
 \end{equation}
In contrast, the precessional motion of the spin polarization is governed solely by the spin torque,
\begin{equation}
\partial_t \hat{\bd{m}}_i  = \bd{\omega}_i \times \hat{\bd{m}}_i .
\end{equation}

In summary, the renormalized precession frequency associated with the angular spin current is
 \begin{eqnarray}
 \bd{\omega}_i & = & - \bd{h}_i -  \sum_{j} \frac{\hat{\bd{m}}_i}{ | \bd{m}_i | } \times
 \langle \bd{I}_{ i \rightarrow j } \rangle,
 \label{eq:omegageneral}
 \end{eqnarray}
while the operator representing the translational spin current is
 \begin{eqnarray}
 {I}^{\parallel}_{ i \rightarrow j} & = & \hat{\bd{m}}_i \cdot   \bd{I}_{i \rightarrow j} .
 \label{eq:Igeneral}
 \end{eqnarray}
Note that  time-dependent
changes in the length of the magnetization 
are always accompanied by translational spin transport.
On the other hand, stationary translational spin currents are also possible
if the length of the magnetization is constant \cite{Schuetz03}.

\subsection{Localized spins}

The above expressions have been derived for a lattice model for itinerant electrons.
It is instructive to work out the explicit form of the rotation vector $\bd{\omega}_i$ and
the translational spin current operator $I^{\parallel}_{ i \rightarrow j }$ 
for a localized spin model containing only the spin degrees of freedom.
For simplicity, let us specify the interaction to the on-site Hubbard interaction
${\cal{U}} = U \sum_i n_{i \uparrow} n_{i \downarrow}$, where $n_{i \sigma} = 
 c^{\dagger}_{ i \sigma} c_{ i \sigma}$.  Assuming 
$U \gg | t_{ij}^{\sigma \sigma^{\prime}} |$ and a half-filled lattice, 
we can use a canonical transformation \cite{MacDonald88} to derive from Eq.~(\ref{eq:Hubbard}) an effective Hamiltonian involving only spin $1/2$ operators $\bd{S}_i$
acting on the reduced Hilbert space of singly occupied lattice sites.
The effective spin  Hamiltonian can be written as\cite{Moriya60}
 \begin{align} \label{eq:Hspin}
 {\cal{H}}^{\rm spin}  =
 & -\frac{1}{2U} \sum_{ij} \left( |t_{ij}|^2 + | \bm{\lambda}_{ij} |^2 \right)
 \nonumber\\
 & +
 \frac{1}{2} \sum_{ij} \sum_{\alpha \beta} 
 \mathbb{K}_{ij}^{\alpha \beta} {S}^{\alpha}_i {S}^{\beta}_j 
 - \sum_i \bd{h}_i \cdot \bd{S}_i,
 \end{align}
where the spin-spin interaction tensor has three contributions,
 \begin{equation}
  \mathbb{K}_{ij}^{\alpha \beta} =   \delta_{ \alpha \beta}   J_{ij}
 + \epsilon_{\alpha \beta \gamma} D^{\gamma}_{ij} + \Gamma_{ij}^{\alpha \beta}.
 \label{eq:Jtens}
 \end{equation}
Here the isotropic exchange coupling $J_{ij}$ and the antisymmetric
Dzyaloshinskii-Moriya vector $\bd{D}_{ij}$ are given by
 \begin{align}
 J_{ij} & = \frac{4}{U} \left( |t_{ij}|^2 - | \bm{\lambda}_{ij} |^2 \right) ,
 \\
 \bm{D}_{ij} & = - \frac{8}{U} \textrm{Re} \left[ t_{ij} \bm{\lambda}_{ji} \right]
 = \frac{8}{U} \textrm{Re} \left[ t_{ij} \bm{\lambda}_{ij}^* \right],
 \end{align}
while $\Gamma_{ij}^{\alpha \beta}$ is a symmetric tensor in spin space with
matrix elements
 \begin{equation}
  {\Gamma}_{ij}^{\alpha \beta} 
  = - \frac{4}{U} \left( \lambda_{ij}^\alpha \lambda_{ji}^\beta + \lambda_{ij}^\beta \lambda_{ji}^\alpha \right)
  = \frac{8}{U} \textrm{Re} \left[ \lambda_{ij}^\alpha ( \lambda_{ij}^\beta )^* \right] .
 \end{equation}
The Heisenberg equation of motion can be written as
 \begin{equation}
   \partial_t \bd{S}_i + \sum_j {\bd{I}}^{\rm spin}_{ i \rightarrow j} = - \bd{h}_i \times
 {\bd{S}}_i,
 \end{equation}
where the operator
 \begin{equation} 
 {\bd{I}}^{\rm spin}_{ i \rightarrow j} = {\bd{S}}_i \times {\mathbb{K}}_{ij} \bd{S}_j
 \label{eq:Istrong}
 \end{equation}
is the strong coupling limit of the operator $\bd{I}_{ i \rightarrow j }$ 
defined in Eq.~(\ref{eq:Islattice})
in the
reduced spin Hilbert space.
Here $\mathbb{K}_{ij}$ is a tensor in spin space with matrix elements given by
Eq.~(\ref{eq:Jtens}).
Alternatively, Eq.~(\ref{eq:Istrong}) can be obtained directly from 
Eq.~(\ref{eq:Islattice}) via a canonical transformation \cite{MacDonald88}.
With the substitution $\bd{I}_{ i \rightarrow j } \rightarrow \bd{I}^{\rm spin}_{ i \rightarrow j } $
the expressions (\ref{eq:omegageneral}) and (\ref{eq:Igeneral})
for the local precession frequency and the longitudinal spin transport operator
remain valid, so that we obtain
\begin{align}
 \bd{\omega}_i = 
 & - \bd{h}_i -  \sum_{j} \frac{\hat{\bd{m}}_i}{ | \bd{m}_i | } \times
 \langle    {\bd{S}}_i \times {\mathbb{K}}_{ij} \bd{S}_j    \rangle,
 \label{eq:omegaspin}
 \\
 \langle {I}^{\parallel}_{ i \rightarrow j} \rangle = 
 & \hat{\bd{m}}_i \cdot  \langle    {\bd{S}}_i \times {\mathbb{K}}_{ij} \bd{S}_j    \rangle   .
 \label{eq:Ispinparallel}
 \end{align}
In the classical limit and at vanishing temperature
the quantum mechanical expectation value 
$\langle    {\bd{S}}_i \times {\mathbb{K}}_{ij} \bd{S}_j \rangle $
can be factorized,
 \begin{equation}
 \langle    {\bd{S}}_i \times {\mathbb{K}}_{ij} \bd{S}_j    \rangle \rightarrow
\langle    {\bd{S}}_i \rangle \times {\mathbb{K}}_{ij}   \langle  \bd{S}_j    \rangle
\equiv
 \bd{m}_i \times {\mathbb{K}}_{ij} \bd{m}_j .
 \end{equation}
Consequently the expectation value of the longitudinal spin current operator vanishes
identically in this limit.
This implies that for a generic magnetic insulator 
with a spin Hamiltonian of the form of Eq. (\ref{eq:Hspin})
incoherent thermal or quantum fluctuations are a necessary prerequisite for the transport of actual magnetization.
The local precession frequency reduces in the same limit to
  \begin{align}
  \bd{\omega}_i 
	= & - \bd{h}_i - \sum_j \hat{\bd{m}}_i \times ( \hat{\bd{m}}_i \times
 {\mathbb{K}}_{ij} \bd{m}_j) 
 \nonumber
 \\
 = &  - \bd{h}_i + \sum_{j}  [ {\mathbb{K}}_{ij} \bd{m}_j - 
 \hat{\bd{m}}_i ( \hat{\bd{m}}_i \cdot {\mathbb{K}}_{ij} \bd{m}_j ) ].
 \label{eq:omegaires}
 \end{align}
The last term in Eq.~(\ref{eq:omegaires}) is proportional to $\hat{\bm{m}}_i$
and hence does not contribute to
$\bd{\omega}_i \times \bd{m}_i$ so that we may write 
$\bd{\omega}_i  \times \bd{m}_i = 
 ( - \bd{h}_i - \delta \bd{h}_i ) \times \bd{m}_i $, where the 
renormalization of the magnetic field is given by
 \begin{equation}
 \delta \bd{h}_i  = - \sum_{j}   {\mathbb{K}}_{ij} \bd{m}_j.
 \label{eq:hex2}
 \end{equation}
Note that Eq.~(\ref{eq:hex2}) can also
be obtained by means of a simple mean-field decoupling
of the spin Hamiltonian in Eq.~(\ref{eq:Hspin}).

\section{Spin superfluidity and persistent spin currents}
\label{sec:supercurrents}

To illustrate the differences between translational and angular spin transport
and how this affects the interpretation of spin supercurrents,
it is instructive to consider simple model systems
that can support translational and angular spin currents in
equilibrium or in metastable states.

\subsection{Easy-plane ferromagnet}

Let us first consider an easy-plane ferromagnet
described by the spin $S$ Heisenberg Hamiltonian
\begin{equation}
{\cal H}^{\textrm{plane}} = -\frac{1}{2} \sum_{ij} J_{ij} \bm{S}_i\cdot\bm{S}_j
+\frac{K}{2} \sum_i S_i^z S_i^z
\end{equation}
on a simple cubic lattice,
with exchange coupling $J_{ij}=J>0$ for nearest neighbors only, 
and an easy-plane anisotropy $K>0$.
This kind of systems has served in the literature as elementary example for 
spin superfluidity \cite{MacDonald01,Sonin10,Sonin13,Takei14a,Flebus16,Chen16}.
We will now calculate the local precession frequency (\ref{eq:omegaspin}) and the
translational spin current (\ref{eq:Ispinparallel})
for this system to leading order in an $1/S$ expansion.
To facilitate this we expand the spin operators in a local basis
defined by the instantaneous direction of the spin polarization 
$ \hat{\bm{m}}_i(t) = \langle \bm{S}_i(t) \rangle / | \langle \bm{S}_i(t) \rangle | $:
\begin{equation}
\bm{S}_i = S_i^\parallel \hat{\bm{m}}_i 
+ S_i^{(1)} \bm{e}_i^{(1)} + S_i^{(2)} \bm{e}_i^{(2)} .
\end{equation}
Here,
$ \bm{e}_i^{(1)}(t) $ and $ \bm{e}_i^{(2)}(t) $
are unit vectors chosen such that
$ \{ \bm{e}_i^{(1)}, \bm{e}_i^{(2)}, \hat{\bm{m}}_i \} $
form a right-handed basis at every lattice site.
In this basis we may now bosonize the spin operators by
means of a Holstein-Primakoff (HP) transformation,
\begin{subequations} \label{eq:HP}
\begin{align}
S_i^\parallel =& S- a_i^\dagger a_i , \\
S_i^{(1)} + i S_i^{(2)} =&  \sqrt{2S} a_i + {\cal O}(S^{-1/2}) ,
\end{align}
\end{subequations}
where the $a_i$ are canonical Bose operators.
Especially note that since we self-consistently define the quantization axis as
the direction of the local magnetization,
$ \hat{\bm{m}}_i(t) = \langle \bm{S}_i(t) \rangle / | \langle \bm{S}_i(t) \rangle | $,
by definition the HP bosons can never condense \cite{Rueckriegel12}.
This is completely analogous to the fact that in the superfluid phase of interacting bosons 
the Bogoliubov quasi-particles, 
which are the Goldstone modes
associated with the spontaneous breaking of the U(1)-symmetry in the superfluid state, 
do not condense provided the condensate wave-function is self-consistently
defined via the solution of the Gross-Pitaevskii equation.

\begin{figure}[tb]
\begin{center}
 \includegraphics[width=0.4\textwidth]{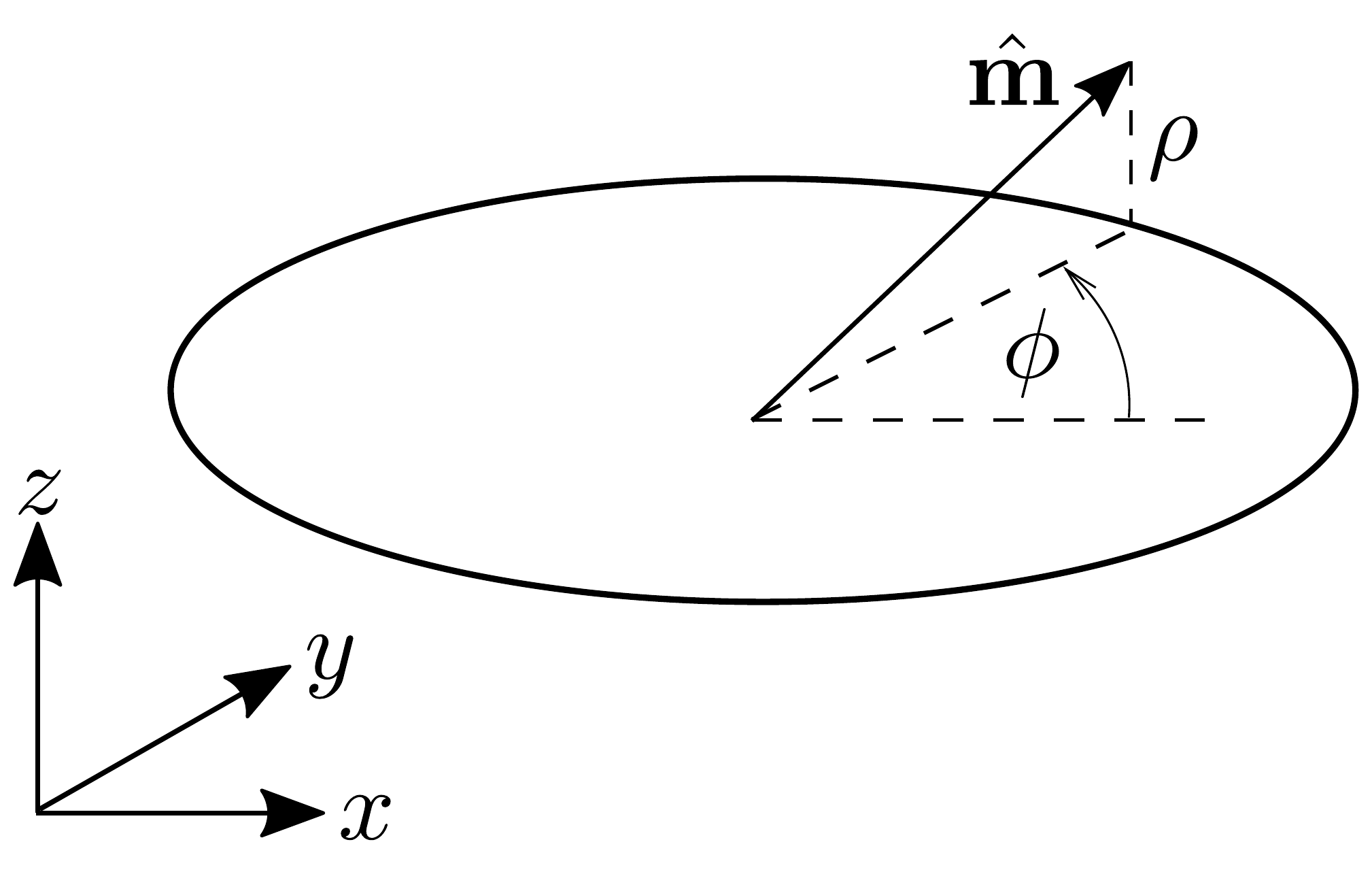}
\end{center}
\caption{%
Magnetic polarization in the easy-plane ferromagnet.
$\rho$ denotes the deviation from the ground state with the magnetization lying in the $x$-$y$-plane.
Superfluid angular transport is possible due to the $U(1)$ freedom of the phase $\phi$ in this plane.
}
\label{fig:easy_plane}
\end{figure}

With the HP bosonization (\ref{eq:HP}), 
we find that the leading order contributions to 
the precession frequency and the spin current are of order $S$.
Explicitly, the local precession frequency (\ref{eq:omegaspin}) 
becomes  
\begin{equation} \label{eq:omegaS}
\bm{\omega}_i = -\sum_j SJ_{ij} \hat{\bm{m}}_j + SK (\hat{\bm{m}}_i\cdot\bm{e}_z) \bm{e}_z .
\end{equation}
To this order, the polarization equation of motion (\ref{eq:polarization})
recasts the Landau-Lifshitz equation of classical spin dynamics.
Assuming that the magnetic texture $\hat{\bm{m}}_i=\hat{\bm{m}}(\bm{R}_i)$ varies
only slowly in space,
we can take the continuum limit.
The exchange contribution to the polarization equation of motion 
can then be identified with the divergence of the classical spin current,
\begin{equation}
\sum_j SJ_{ij} \hat{\bm{m}}_j \times \hat{\bm{m}}_i \to \sum_\mu \partial_\mu \bm{J}^\mu ,
\end{equation}
where the classical spin current is defined as
\begin{equation} \label{eq:Jclassical}
\bm{J}^\mu = - SJa^2 \hat{\bm{m}} \times \partial_\mu \hat{\bm{m}} .
\end{equation}
Here $\partial_\mu = \partial / \partial r^\mu$, and $a$ is the distance between nearest neighbors.
An exact equilibrium solution for the polarization is then given by
$ \bm{\hat{m}}(\bm{r}) = 
\left[ \bm{e}_x \cos\phi(\bm{r}) + \bm{e}_y \sin\phi(\bm{r}) \right] $,
with a local phase satisfying $\nabla^2 \phi(\bm{r})=0$.
The $U(1)$ freedom associated with the choice of $\phi(\bm{r})$ lies at 
the heart of the concept of spin superfluidity \cite{Sonin10,Chen16}.
Out of equilibrium we can make the general ansatz
\begin{align}
\bm{\hat{m}}(\bm{r},t) =
& \sqrt{1-\rho^2(\bm{r},t)}
\left[ \bm{e}_x \cos\phi(\bm{r},t) + \bm{e}_y \sin\phi(\bm{r},t) \right]
\nonumber\\[.2cm]
&
+ \bm{e}_z \rho(\bm{r},t) ,
\end{align}
which is depicted graphically in Fig.~\ref{fig:easy_plane}.
If the system is only slightly driven out of equilibrium,
the equation of motion (\ref{eq:polarization}) for the spin polarization
becomes to lowest nonvanishing order in deviations from equilibrium
\begin{subequations} \label{eq:Josephson}
\begin{align}
\partial_t \phi =& SK \rho , \\
\partial_t \rho =& SJa^2 \nabla^2 \phi .
\end{align}
\end{subequations}
The above Eqs. (\ref{eq:Josephson}) are of the form of the Josephson equations of superconductivity;
hence they are conventionally interpreted as describing a spin supercurrent carried by the magnetization texture \cite{Sonin10,Takei14a}.
However, since this current does not correspond to translational transport of the
local magnetization $|\bm{m}_i|=|\langle \bm{S}_i \rangle |$,
this supercurrent is \emph{not} equivalent to the superfluid transport of magnetic moments.
In particular, 
a stationary angular supercurrent with $\nabla\phi=\textrm{const}$ is simply an
inhomogeneous magnetic texture;
while it can be viewed as a stationary transport of polarization,
it does not correspond to any physical movement of magnetic moments
which would generate an electrical field \cite{Hirsch99}.

To evaluate the ${\cal O}(S)$ contribution to the translational spin current (\ref{eq:Ispinparallel}),
we will assume for simplicity a slowly varying magnetic texture $\hat{\bm{m}}_i$ in a
metastable superfluid state with $\rho\ll 1$. 
%
%
Applying the HP transformation (\ref{eq:HP}) to the spin current (\ref{eq:Ispinparallel}) then yields
\begin{equation}
\langle I_{i\to j}^\parallel \rangle = 
\textrm{Im} \left[ -2SJ_{ij} \langle a_i^\dagger a_j \rangle
+ \delta_{ij} SK \langle a_i a_i \rangle \right] ,
\end{equation}
which can to be evaluated to
\begin{equation} \label{eq:Iplane}
\langle I_{i\to j}^\parallel \rangle =
\frac{1}{Na} \sum_{\bm{k}} v_{\bm{k}}^\mu \left[
n_{\bm{k}} + \frac{1}{2} \left( 1 - \frac{E_{\bm{k}}}{A_{\bm{k}}} \right)
\right] ,
\end{equation}
where $\bm{R}_i$ and $\bm{R}_j=\bm{R}_i+a\bm{e}_\mu$ are nearest neighbor lattice sites.
Here $n_{\bm{k}}$ is the distribution functions of magnons with dispersion
\begin{equation} \label{eq:Ek}
E_{\bm{k}} = S \sqrt{ \left( K + J_{\bm{k}=0} - J_{\bm{k}} \right) 
\left( J_{\bm{k}=0} - J_{\bm{k}} \right) } 
\end{equation}
and velocity 
$ v_{\bm{k}}^\mu = \partial E_{\bm{k}} / \partial k^\mu $,
and the remaining coefficient of the quantum-mechanical zero-point fluctuations is
\begin{equation}
A_{\bm{k}} = S\left( J_{\bm{k}=0} - J_{\bm{k}} \right) + SK/2 .
\end{equation}
A detailed derivation of the Hamiltonian in the local basis 
and of the magnon dispersion (\ref{eq:Ek}) is relegated to the Appendix.
In equilibrium the translational current (\ref{eq:Iplane}) of course vanishes by symmetry,
i.e.,
$ \langle I_{i\to j}^\parallel \rangle = 0 $.
Therefore there is no translational movement of magnetic moment 
in this superfluid spin state. 

Lastly,
let us note that the complete decoupling of the magnetic texture and the incoherent magnons 
is an artifact of the lowest order approximation in the $1/S$ expansion,
and of the assumption of a slowly varying texture.
If one relaxes either of these approximations,
there will be a coupling,
resulting in a two-fluid description of spin transport
like the one discussed in [\onlinecite{Flebus16}].
We emphasize however that from the point of view we adopted,
the superfluid does not arise due to the condensation of magnons.
We rather consider the magnons as fluctuations on top of the superfluid ground state;
these magnons cannot condense by definition.

\subsection{Heisenberg ring}

\begin{figure}[tb]
\begin{center}
 \includegraphics[width=0.4\textwidth]{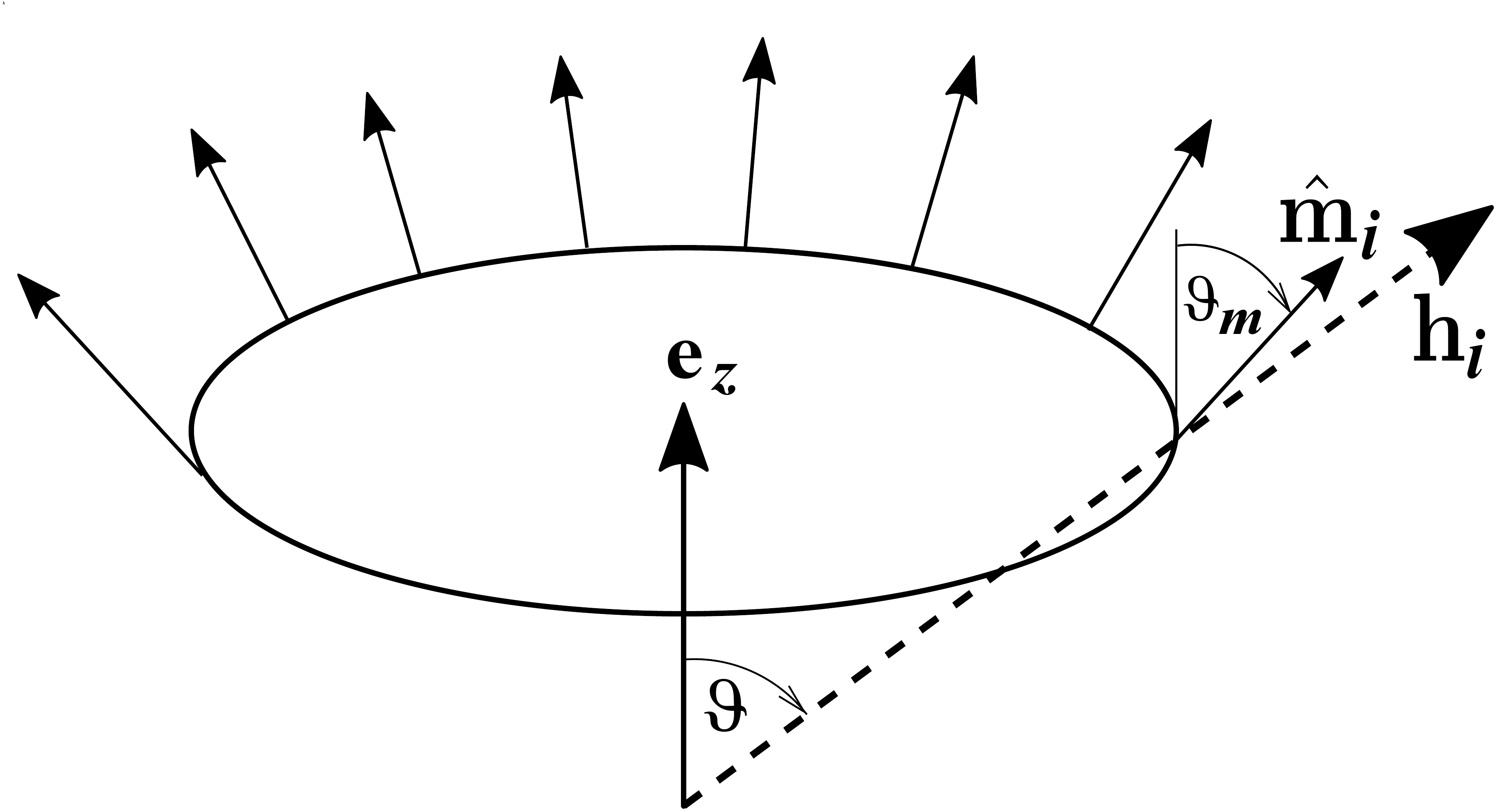}
\end{center}
\caption{%
Magnetization configuration of a ferromagnetic  ring in a radial magnetic field.
Due to the exchange field the angle $\vartheta_m$ between the magnetization and the
$z$-axis is slightly smaller then the angle $\vartheta$ between the
external field and the $z$-axis, see Ref.~[\onlinecite{Schuetz03}].
}
\label{fig:crown}
\end{figure}

Next, 
let us consider a system where the translational spin current is finite even in equilibrium,
i.e., 
there is a persistent translational spin current corresponding to the phyiscal movement of magnetic moments.
Consider a ferromagnetic spin $S$ Heisenberg model  in a radial
inhomogeneous magnetic field with quantum mechanical Hamiltonian
 \begin{equation}
 {\cal{H}}^{\textrm{ring}} = -\frac{1}{2} \sum_{ ij} J_{ij} \bd{S}_i \cdot \bd{S}_j - \sum_{i} \bd{h}_i
 \cdot \bd{S}_i,
 \label{eq:hamiltoniancrown}
 \end{equation}
where the sums are over the $N$  sites of the lattice of localized  
spins on the ring coupled by ferromagnetic exchange interactions
$J_{ij} = J>0$ if $i$ and $j$ label nearest neighbors.
$\bd{h}_i$ is a crown-shaped
inhomogeneous magnetic field of the form
 \begin{equation} \label{eq:hCrown}
 \bd{h}_i = h [  
  \sin \vartheta (  \bd{e}_x \cos \varphi_i + \bd{e}_y \sin \varphi_i  ) 
	+ \bd{e}_z \cos \vartheta ],
 \end{equation}
where the angles $\varphi_i$ label the positions of the spins on the ring,
as illustrated in Fig.~\ref{fig:crown}.
In Ref.~[\onlinecite{Schuetz03}] it has been shown that 
at finite temperature $T > 0$ spin-wave excitations carry a persistent 
equilibrium spin current
\begin{equation} \label{eq:Iring}
\langle I_{i\to i+1}^\parallel \rangle =
\frac{1}{L} \sum_n \frac{v_n}{ e^{(\epsilon_n + |\bm{h}|)/T} - 1 }
\end{equation}
circulating the ring.
Here $L$ is the length of the ring,
$v_n=\partial \epsilon_n / \partial k_n$ is the magnon velocity, 
and the magnon dispersion is 
$\epsilon_n=SJa^2 k_n^2$,
with lattice spacing $a$ 
and quantized wavevectors 
$k_n=\frac{2\pi}{L}\left( n - \frac{\Omega}{2\pi} \right)$.
$\Omega$ is the solid angle traced out by the local magnetization direction $\hat{\bm{m}}_i$
on the unit sphere in order parameter space as it moves around the chain.
This finite solid angle $\Omega$, i.e., 
the topology of the spin configuration on the ring,
is responsible for the finiteness of the equilibrium current (\ref{eq:Iring}).
The situation is completely analogous to persistent electrical currents
in mesoscopic metal rings pierced by a magnetic flux \cite{Schuetz03}.
These electrical currents generate a magnetic dipole field;
one of us has shown in Ref.~[\onlinecite{Schuetz03}] 
that the persistent spin current (\ref{eq:Iring}) similarly generates 
an electric dipole field.
This can be understood as follows \cite{Hirsch99}:
A magnetic dipole moment $\bm{m}$ moving with velocity $\bm{v}$
generates a magnetic field $\bm{B}$ in its rest frame.
Lorentz-transforming back to the laboratory frame,
we find that to lowest order in $\bm{v}/c$ (where $c$ is the speed of light)
this magnetic field generates an electrical field
$ \bm{E} = - \frac{\bm{v}}{c} \times \bm{B} $.

At zero temperature the equilibrium current (\ref{eq:Iring}) vanishes because there
are no spin waves in the ferromagnetic ground state.
However, Bruno and Dugaev \cite{Bruno05}
pointed out that in this system the classical spin current 
$\bm{J}^\mu $ 
defined in Eq.~(\ref{eq:Jclassical}) 
is finite and argued that
therefore the system exhibits an equilibrium spin supercurrent even at $T=0$.
While this interpretation is possible,
we stress that this classical current is angular and not translational;
hence it cannot be associated
with the motion of magnetic dipoles,
but should rather be considered as a renormalization of the external magnetic field \cite{Schuetz04}.
To understand this, let us explicitly calculate the
inhomogeneous magnetization configuration in the classical ground state
of the Hamiltonian (\ref{eq:hamiltoniancrown}).
Just as in the easy-plane ferromagnet, 
the equation of motion for the magnetic texture $\hat{\bm{m}}_i$
is to leading order in $1/S$ the classical Landau Lifshitz equation
\begin{equation} \label{eq:LLring}
\partial_t \hat{\bm{m}}_i = \hat{\bm{m}}_i \times ( \bm{h}_i + \delta \bm{h}_i^{\textrm{ex}} ) ,
\end{equation}
where the exchange field is 
$ \delta \bm{h}_i^{\textrm{ex}} = S\sum_{ij} J_{ij} \hat{\bm{m}}_j $.
The equilibrium solution of the Landau-Lifshitz equation (\ref{eq:LLring}) is
\begin{equation} \label{eq:GroundStateRing}
\hat{\bd{m}}_i = 
   \sin \vartheta_m( \bd{e}_x\cos \varphi_i  + \bd{e}_y \sin \varphi_i )
	 + \bd{e}_z \cos \vartheta_m ,
\end{equation}
where the angle $\vartheta_m$ is slightly smaller than the angle $\vartheta$ 
between the magnetic field and the $z$-axis, 
as discussed in Ref.~[\onlinecite{Schuetz03}].
The deviation of $\vartheta_m$ from $\vartheta$ is determined by
the exchange field $\delta\bm{h}_i^{\textrm{ex}}$.
As for the easy-plane ferromagnet discussed in the last section,
the continuum limit of the exchange torque is the divergence of the classical spin current,
$ \delta\bm{h}_i^{\textrm{ex}} \times \hat{\bm{m}}_i \to \sum_\mu \partial_\mu \bm{J}^\mu $.
For the classical ground state (\ref{eq:GroundStateRing}),
the classical spin current has a finite component 
$\propto \bm{e}_z \nabla \varphi$.
However, as in the easy-plane ferromagnet,
this finite current merely signals an inhomogeneous magnetization configuration
and is not related to the physical transport of magnetization;
hence it also will not generate any electrical field.
If one would on the other hand incorrectly associate 
the classical spin current $\bm{J}^\mu$ with the stationary movement of physical dipoles,
it would have to be accompanied by an electrical field.
This would imply that a purely static inhomogeneous 
magnetization configuration is always accompanied by an electric field, 
in contradiction with the elementary fact 
that in classical electromagnetism magnetostatics and electrostatics are completely decoupled.

Finally, 
let us also point out that although the classical ground state (\ref{eq:GroundStateRing}) of the Heisenberg ring is very similar to the classical ground state of the easy-plane ferromagnet
discussed in the last section,
it does not support spin superfluidity, i.e., angular spin supercurrents,
because the phase $\varphi_i$ of the magnetization 
is pinned by the external magnetic field $\bm{h}_i$, Eq. (\ref{eq:hCrown}).


\section{Summary and conclusions}
\label{sec:conclusions}

We conclude that in the classical limit the angular spin current
introduced by Sun and Xie \cite{Sun05}
can be absorbed into a renormalization of the external magnetic field 
which contributes to the torque acting on the magnetic moments. 
In equilibrium, this term does not describe any current of magnetic moments because
the equilibrium configuration of the magnetization is such that
the total torque on each moment vanishes.
Translational transport of spins is described by the longitudinal spin current defined in Eq. (\ref{eq:Ispinparallel}),
which is only finite due to thermal or quantum fluctuations.
An example for a system exhibiting a finite longitudinal spin current
in equilibrium is a mesoscopic Heisenberg ring in a crown-shaped magnetic field, as discussed
in Ref.~[\onlinecite{Schuetz03}].

Our considerations imply that in equilibrium the classical spin current
$\bm{J}^{\mu}$ 
defined in Eq.~(\ref{eq:Jclassical}) 
does not describe any motion of magnetization.
While it can be interpreted as a stationary current of magnetic polarization,
such an interpretation is by no means mandatory since there is no phyiscal movement.
In nonequilibrium on the other hand,
the classical spin current $\bm{J}^{\mu}$ contributes to the angular spin current,
i.e., to the precessional motion of the spins,
and does transport spin polarization.
In particular, 
this implies that spin superfluidity, 
which is based on the formal similarity  of Eq.~(\ref{eq:Josephson})
with a mass supercurrent of superfluid  bosons \cite{Sonin10,Chen16},
does not correspond to the physical movement of \textit{magnetization}, but of \textit{polarization}.
This means that in equilibrium,
a superfluid spin state will not be accompanied by an electrical field,
in contrast to a persistent translational spin current \cite{Schuetz03}.
This physical difference between the two types of spin transport persists also out of equilibrium:
As already shown by Sun and Xie \cite{Sun05},
an angular spin current with finite $\bm{\omega}\times\hat{\bm{m}}$
will generate an electrical field $ \bm{E}^\omega \propto 1/r^2 $ for large distance $r$ from the source,
whereas the electrical field of translational spin currents decays as $ \bm{E}^\parallel \propto 1/r^3 $.

We have also shown that spin superfluidity can be described entirely 
without referring to off-diagonal long-range order \cite{Yang62,Kohn70} and magnon condensation.
This is achieved by quantizing the spins in a self-consistently defined frame of reference
with the local $z$-axis pointing in the direction $\hat{\bm{m}}$ of the instantaneous magnetization.
Magnons defined with respect to this reference frame can never condense or display off-diagonal long-range order,
hence they are not superfluid.
This is in agreement with the general proof of Kohn and Sherrington \cite{Kohn70}
that bosonic quasi-particles 
which are formed as bound states of particle-hole pairs of the underlying fermionic system 
(such as excitons or magnons) 
do not exhibit off-diagonal long-range order in coordinate space.
Hence, the Bose-Einstein condensation of this type of bosons is not accompanied by superfluidity.
Although the change in magnetic order in a magnetic insulator can be viewed as Bose-Einstein condensation of magnons \cite{Demokritov06,Rueckriegel15,Flebus16,Chen16}, 
the resulting state can always be characterized by magnons that exhibit neither off-diagonal long-range order nor superfluidity.

\section*{ACKNOWLEDGMENTS} 
This work was financially supported by the DFG via SFB/TRR49.
Our ideas on spin transport were sharpened during a workshop on
\textit{Quantum Spintronics: Spin Transport Through Quantum Magnetic Materials}
at the Spin Phenomena Interdisciplinary Center (SPICE) at the University of Mainz, Germany.

\begin{appendix}
\renewcommand{\theequation}{A\arabic{equation}}

\section*{APPENDIX: MAGNONS IN THE ROTATING REFERENCE FRAME}
\setcounter{equation}{0}
\renewcommand{\theequation}{A\arabic{equation}}



This Appendix is devoted to the derivation of the dispersion (\ref{eq:Ek})
of the magnons in the local reference frame defined by the magnetic polarization 
$ \hat{\bm{m}}_i(t) = \langle \bm{S}_i(t) \rangle / | \langle \bm{S}_i(t) \rangle | $ .
We have already derived the general setup of a spin-wave expansion in this local and possibly time dependent reference frame
in Ref.~[\onlinecite{Rueckriegel12}].
The first step is to rotate the $z$-axis of the laboratory frame to the direction $\hat{\bm{m}}_i(t)$ of the local magnetization
by means of an unitary transformation
$ {\cal U}(t) $
acting on the spins.
The explicit form of 
$ {\cal U}(t) $
is given in Ref.~[\onlinecite{Rueckriegel12}].
It then turns out that the rotated Hamiltonian
$
\tilde{{\cal H}}^{\textrm{plane}} = {\cal H}^{\textrm{plane}} + {\cal H}_B
$
contains an additional Berry-phase term acting as a magnetic field due to the time dependence of $\hat{\bm{m}}_i(t)$,
\begin{equation}
{\cal H}_B = -i {\cal U}^\dagger \partial_t {\cal U}
= - \sum_i \bm{B}_i \cdot \bm{S}_i .
\end{equation}
Writing the spin polarization as
\begin{equation}
\hat{\bm{m}}_i = \sin\theta_i 
\left[ \bm{e}_x \cos\phi_i + \bm{e}_y \sin\phi_i \right]
+ \bm{e}_z \cos\theta_i ,
\end{equation}
we can choose the transverse basis vectors as
\begin{align}
\bm{e}_i^{(1)} =& - \bm{e}_x \sin\phi_i + \bm{e}_y \cos\phi_i , \\
\bm{e}_i^{(2)} =& -\cos\theta_i 
\left[ \bm{e}_x \cos\phi_i + \bm{e}_y \sin\phi_i \right]
+ \bm{e}_z \sin\theta_i .
\end{align}
Expanding the Berry-phase magnetic field in this basis,
$ \bm{B}_i = B_i^{(1)} \bm{e}_i^{(1)} + B_i^{(2)} \bm{e}_i^{(3)} + B_i^{\parallel} \hat{\bm{m}}_i $,
we explicitly find
\begin{subequations}
\begin{align}
B_i^{(1)} =& \partial_t \theta_i
= - \bm{e}_i^{(2)} \cdot \partial_t \hat{\bm{m}}_i , \\
B_i^{(2)} =& \sin \theta_i \partial_t \phi_i
= \bm{e}_i^{(1)} \cdot \partial_t \hat{\bm{m}}_i , \\
B_i^{\parallel} =& \cos\theta_i \partial_t \phi_i .
\end{align}
\end{subequations}
Applying the Holstein-Primakoff transformation (\ref{eq:HP})
%
%
%
%
to the rotated Hamiltonian then yields
\begin{equation}
\tilde{{\cal H}}^{\textrm{plane}} = E_0 + {\cal H}_1 + {\cal H}_2 + {\cal O}(S^{1/2}) .
\end{equation}
Here the classical ground state energy is
\begin{equation}
E_0 = -\frac{S^2}{2} \sum_{ij} J_{ij} \hat{\bm{m}}_i\cdot\hat{\bm{m}}_j
+\frac{S^2}{2} K \sum_i (\hat{\bm{m}}_i\cdot\bm{e}_z)^2 
- S \sum_i B_i^\parallel .
\end{equation}
The term linear in the Bose operators can be written as
\begin{equation}
{\cal H}_1 =
\sqrt{\frac{S}{2}} \sum_i 
a_i \left( \bm{e}_i^{(2)} +i \bm{e}_i^{(1)} \right) \cdot
   \left(  \partial_t \hat{\bm{m}}_i - \bm{\omega_i} \times \hat{\bm{m}}_i \right) 
 + \textrm{h.c.} ,
\end{equation}
where $\bm{\omega}_i$ is the ${\cal O}(S)$ local precession frequency, Eq. (\ref{eq:omegaS}). 
Because the magnetic polarization satisfies the equation of motion
$ \partial_t \hat{\bm{m}}_i = \bm{\omega}_i \times \hat{\bm{m}}_i $
up to this order in $1/S$,
we conclude that 
$ {\cal H}_1 $
vanishes identically.
Note that this implies that the Holstein-Primakoff bosons cannot condense,
a statement which remains true to all orders in $1/S$ 
due to the self-consistency of the basis.
Up to this point the discussion is completely general and applies to all magnetic insulators.

To evaluate the quadratic part of the rotated Hamiltonian
we will assume as in the main text that the magnetic texture varies only slowly in space
and is only slightly out of equilibrium.
Then we may approximate
$ \hat{\bm{m}}_i \cdot \hat{\bm{m}}_{i+1} \approx 1 $,
$ \bm{e}_i^{(p)} \cdot \bm{e}_{i+1}^{(p')} \approx \delta^{pp'} $,
$ \bm{e}_i^{(p)} \cdot \hat{\bm{m}}_{i+1} \approx 0 $,
with $p,p'\in\{1,2\}$,
and
$ \cos\theta_i \approx 0 $.
With this simplifications we immediately obtain
\begin{align}
{\cal H}_2 =
& \sum_{\bm{k}} \Biggl[ 
S\left( J_{\bm{k}=0} - J_{\bm{k}} + \frac{K}{2} \right) a_{\bm{k}}^\dagger a_{\bm{k}}
\nonumber\\
& \phantom{ \sum_{\bm{k}} \Biggl[ }
-S \frac{K}{4} \left( a_{\bm{k}} a_{-\bm{k}} + \textrm{h.c.} \right) + S\frac{K}{4}
\Biggr] ,
\end{align}
which describes free magnons with the dispersion (\ref{eq:Ek}).

\end{appendix}

\end{document}